\documentclass[letter,traditabstract]{aa} 

\usepackage{graphicx}
\usepackage{natbib,twoopt,amssymb,lscape}
\usepackage{lipsum}
\usepackage[breaklinks=true,hidelinks]{hyperref} 
\bibpunct{(}{)}{;}{a}{}{,} 
\newcommandtwoopt{\citeads}[3][][]{\href{http://adsabs.harvard.edu/abs/#3}%
{\citealp[#1][#2]{#3}}} 
\newcommandtwoopt{\citepads}[3][][]{\href{http://adsabs.harvard.edu/abs/#3}%
{\citep[#1][#2]{#3}}} 
\newcommandtwoopt{\citetads}[3][][]{\href{http://adsabs.harvard.edu/abs/#3}%
{\citet[#1][#2]{#3}}} 
\newcommandtwoopt{\citeyearads}[3][][]%
{\href{http://adsabs.harvard.edu/abs/#3}{\citeyear[#1][#2]{#3}}}
\usepackage{graphicx}
\usepackage{longtable}
\usepackage{natbib}
\bibpunct{(}{)}{;}{a}{}{,} 
\usepackage{soulutf8}
\usepackage{txfonts}

\newcommand{\suzaku}{{\it Suzaku}}

\newcommand{\swift}{{\it Swift}}
\newcommand{\integral}{\textit{INTEGRAL}}
\newcommand{\nustar}{\textit{NuSTAR}}
\newcommand{\tess}{\textit{TESS}}

\newcommand{\fluxcgs}{ergs~s$^{-1}$~cm$^{-2}$~\rm}
\newcommand{\lumcgs}{ergs~s$^{-1}$}

\newcommand{\gx}{GX~1+4}

\newcommand{\igr}{IGR~J16194-2810}

\usepackage{graphicx}

\usepackage{lineno}
\usepackage{txfonts}
\begin{document} 

\title{{\it K2} \& \tess\ observations of symbiotic X-ray binaries: \gx\ and \igr }



   \author{G. J. M. Luna \inst{1}}

   \institute{CONICET-Universidad Nacional de Hurlingham, Av. Gdor. Vergara 2222, Villa Tesei, Buenos Aires, Argentina\\
              \email{juan.luna@unahur.edu.ar}
                      }

   \date{Received June 2023; accepted y}

 
  \abstract
   {I analyze the {\it K2} and \tess\ data taken in 2016, 2019 and 2021 of the symbiotic X-ray binaries \gx\ and \igr. \gx\ consists of a pulsar accreting from a red giant companion in a 1160 days orbit. Since 1984, the pulsar has shown a continuous spin-down rate of $\dot{P}$=-0.1177(3) mHZ/yr. I report the detection of the spin period at an average value of 180.426(1) seconds as observed with the {\it K2} mission and confirm that the spin period continues to increase at a rate of $\sim$1.61$\times$10$^{-7}$ s/s. The {\it K2} and hard X-rays, as observed with \swift/{\textrm BAT}, varied in tandem, in agreement with other authors who proposed that the optical light arise from reprocessed X-ray emission.

   In the case of \igr, the X-ray and optical spectroscopy have been interpreted as arising from a neutron star accreting from a M2 III red giant companion. Its orbital period is unknown, while I report here the detection of a modulation with a period of 242.837 min, interpreted as the neutron star spin period. \igr\ is thus the second symbiotic X-ray binary where the spin period is detected in optical wavelengths. This period, however, was only detected during the \tess\ observations of Sector 12 in 2019. The non-detection of this modulation during the observations of Sector 39 in 2021 is perhaps related with the orbital modulation, i.e. a low inclination of the orbit.
   }

   \keywords{binaries: symbiotic, individual: GX 1+4,  IGR J16194-2810  }
\titlerunning{{\it K2} \& \tess\ observations of symbiotic X-ray binaries}
\authorrunning{Luna.}
\maketitle
%

\section{Introduction}
Symbiotic binaries consists of compact object accreting from a red giant companion. Those symbiotics with neutron stars are known as symbiotic X-ray binaries. The current census accounts for about a dozen of these systems \citep{2019MNRAS.485..851Y}. This class of accreting neutron stars is extremely heterogeneous. At first sight, the only feature that these objects share is the presence of an evolved, wind mass-loosing companion, from where the neutron star accretes. Other system parameters such the neutron star spin period, the orbital period, or the accretion luminosity are very different from one system to another. For example, spin periods range from about a 100 s (Sct X-1) to more than 18,000 s (4U~1954+319) \citep[see][]{2019MNRAS.485..851Y}. 

\gx\ (V2116 Oph) was the first member of this class, discovered in X-rays by \citet{1971ApJ...169L..17L} with balloon experiment, obtaining a glimpse of what later would be confirmed as the spin period of about 2 min. The optical counterpart was discovered by \citet{1973NPhS..245...39G} as an M5 III spectral type red giant \citep{1997ApJ...489..254C}. The spin period in optical wavelengths was first reported by \citet{1996IAUC.6489....1J}, and until this study it was the only symbiotic X-ray binary with the spin period detected in optical. The neutron star in \gx\ has since then been identified as an accreting pulsar in a symbiotic binary. 

A long history of the spin behavior of the neutron star in \gx\ exists, with a thoroughly compilation by \citet{2012A&A...537A..66G}. The changes in the spin period of the accreting pulsar are thought to be related with the accretion rate and the torques' changes due to the interaction of the neutron' star magnetic field with the inner and outer region of the truncated accretion disk \citep{1987A&A...183..257W,1979ApJ...234..296G}.

\igr\ was classified as a symbiotic X-ray binary by \citet{2007A&A...470..331M} after the identification of the optical counterpart of the \integral\ hard X-ray source. In their analysis of the \swift/{\textrm XRT} light curve, the authors did not find evidence of pulsation of the neutron star, perhaps because of geometric effects such as low inclination of the binary or close alignment of the rotation and magnetic axis of the neutron star.


In this letter I analyze the exquisite, long term, almost uninterrupted photometric time series of \gx\ and \igr\ obtained by the {\it K2} and \tess\ missions and searched for the neutron star spin period and its possible changes. 
In Section \ref{sec:obs}, I present the data and detail the procedures to extract and remove spurious effects from the light curve and the search for the spin period. Sections \ref{sec:res} and \ref{sec:conc} show and discuss the results.

\section{Observations} \label{sec:obs}

\gx\ was observed during quarter 11 of the {\it K2} mission on 2016 September 24 19:12:30 UT (lc1) and on 2016 October 21 06:17:05 UT (lc2) with a cadence of 1 min during 23.2902 (lc1) and 47.7263 (lc2) days respectively. 
\igr\ was observed with \tess\ during Sectors 12 and 39, starting at 2019-05-21 11:07:37UT and 2021-05-27 06:37:12UT, respectively. During Sector 12 the cadence was 30 min while during Sector 39 it was 10 minutes.

\begin{figure*}
\begin{center}
\includegraphics[scale=0.30]{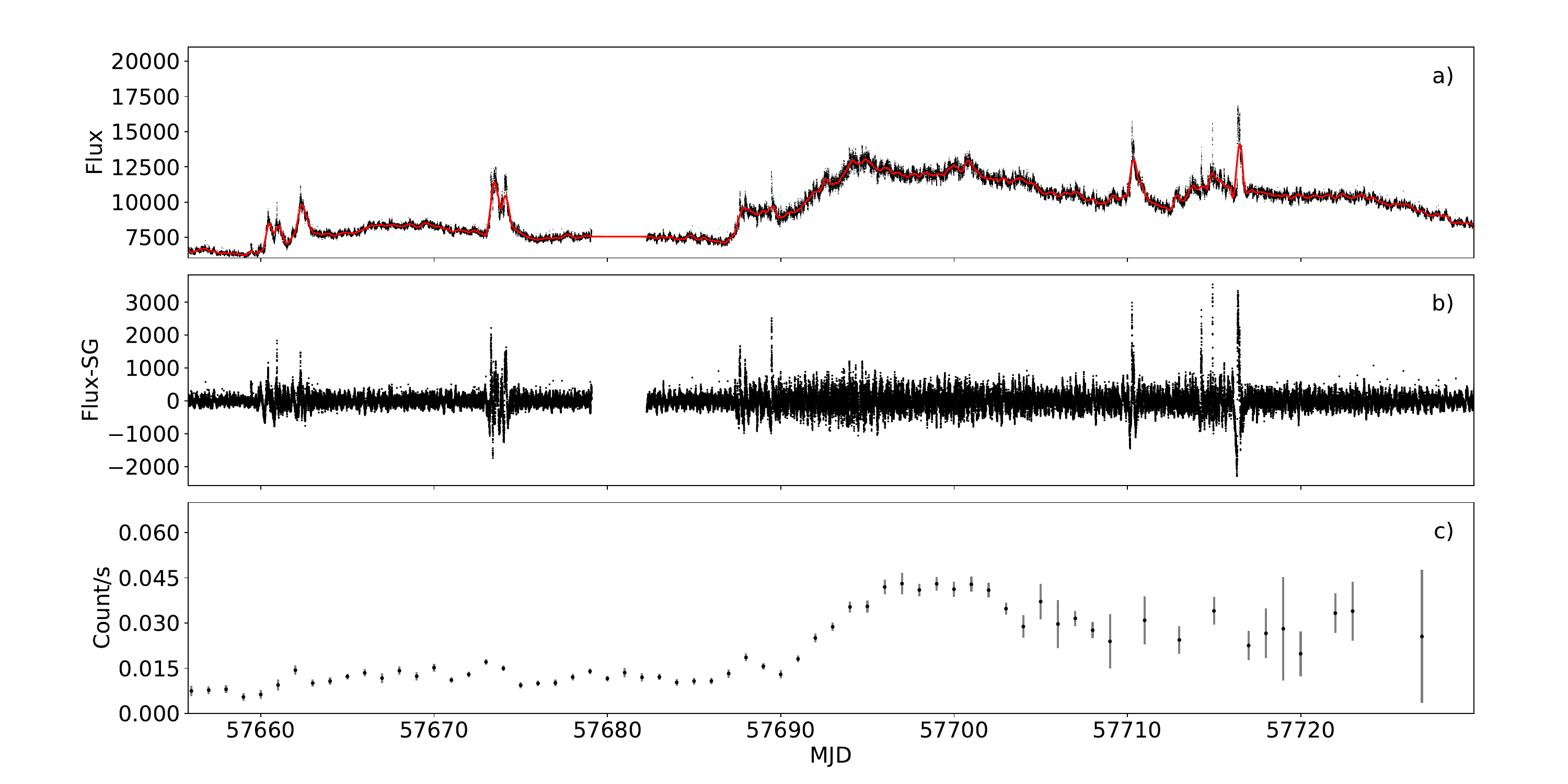}
\caption{{\it a:} \gx\ {\it K2} lc1+lc2 light curve (black dots) with the SG filter overlayed (red line). {\it b}: {\it K2} light curve after subtracting the SG filter- {\it c}: \swift/BAT light curve with 1 day bins in the 15--50 keV energy range. The long term variability in X-rays, from the accretion process, seems to be closely followed by the optical light curve. 
}
\label{bat}
\end{center}
\end{figure*}

\begin{figure*}
\begin{center}
\includegraphics[scale=0.45]{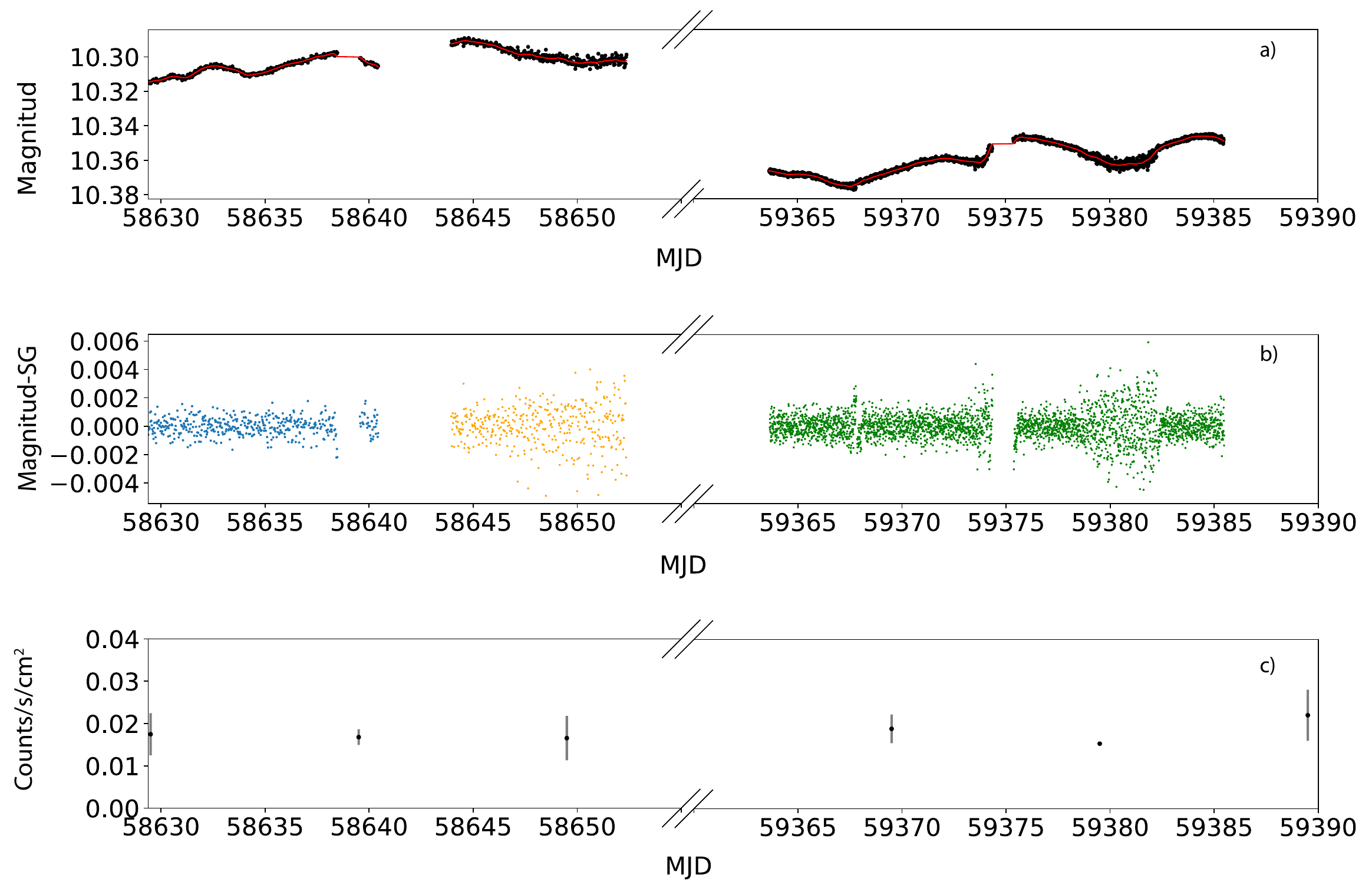}
\caption{{\it a):} \tess\ light curve of \igr\ (black dots) with the SG filter curve on top (red line). {\it b)}: Light curve resulting after the subtraction of the SG filter.
{\it c)}: {\textit MAXI} 2-20 keV light curve with 10-day size bins. }
\label{fig:igr_sg}
\end{center}
\end{figure*}

I used the \texttt{Lightkurve} package \citep{2018ascl.soft12013L} to download the light curves and remove outliers\footnote{as described in \url{https://docs.lightkurve.org/tutorials/index.html}}. \tess\ fluxes (e$^{-}$ s$^{-1}$) were transformed to \tess\ magnitudes using the zero points from \citet{vanderspek2018tess}. 
I then applied a Savitzky-Golay smoothing filter to remove the low frequency variability (Figures \ref{bat} and \ref{fig:igr_sg}). 

In order to search for the presence of the spin period, I used the Generalized Lomb-Scargle (GLS) algorithm as implemented in the \texttt{astropy} library with a "standard"\footnote{The standard normalized periodogram is normalized by the residuals of the data around the constant reference model (see \url{https://docs.astropy.org/en/stable/timeseries/lombscargle.html)}} normalization. Significance levels were calculated by the bootstrapping method implemented on the same library. In the case of the light curve from \gx, I searched for periods around the already known spin period, in the frequency range of 470 to 490 d$^{-1}$. 

The \tess\ light curve of \igr\ presents larger gaps than those from {\it K2} due to satellite downlink and/or bad-quality cadences (conservatively I have only downloaded those measurements with the \texttt{quality\_mask='hard'}). In this case I divided the light curve in three portions: (1) MJD $<$ 58640.43; (2) 58643.97 $>$ MJD $<$ 58652.37 and (3) MJD $>$ 59363.64 (see Figure \ref{fig:igr_sg}) and searched for periods. 

In order to determine the X-ray flux state of \gx\ and \igr, which is related with the accretion state and possibly changes in the spin period, 
I downloaded the \swift/BAT light curve of \gx\ from the \swift/BAT Monitor web\footnote{\url{https://swift.gsfc.nasa.gov/results/transients/}} in the 15--50 keV energy range \citep{2013ApJS..209...14K}, and selected those 1-day bins between the dates observed with {\it K2} (see Figure \ref{bat}). In the case of \igr\, I downloaded the MAXI\footnote{\url{http://maxi.riken.jp/star_data/J1619-281/J1619-281.html}} light curve in the 2--20 keV energy range and rebinned at a 10-days bin size to increase the signal-to-noise ratio.

\section{Results} \label{sec:res}

\subsection{\gx}

\begin{figure*}
\begin{center}
\includegraphics[scale=0.68]{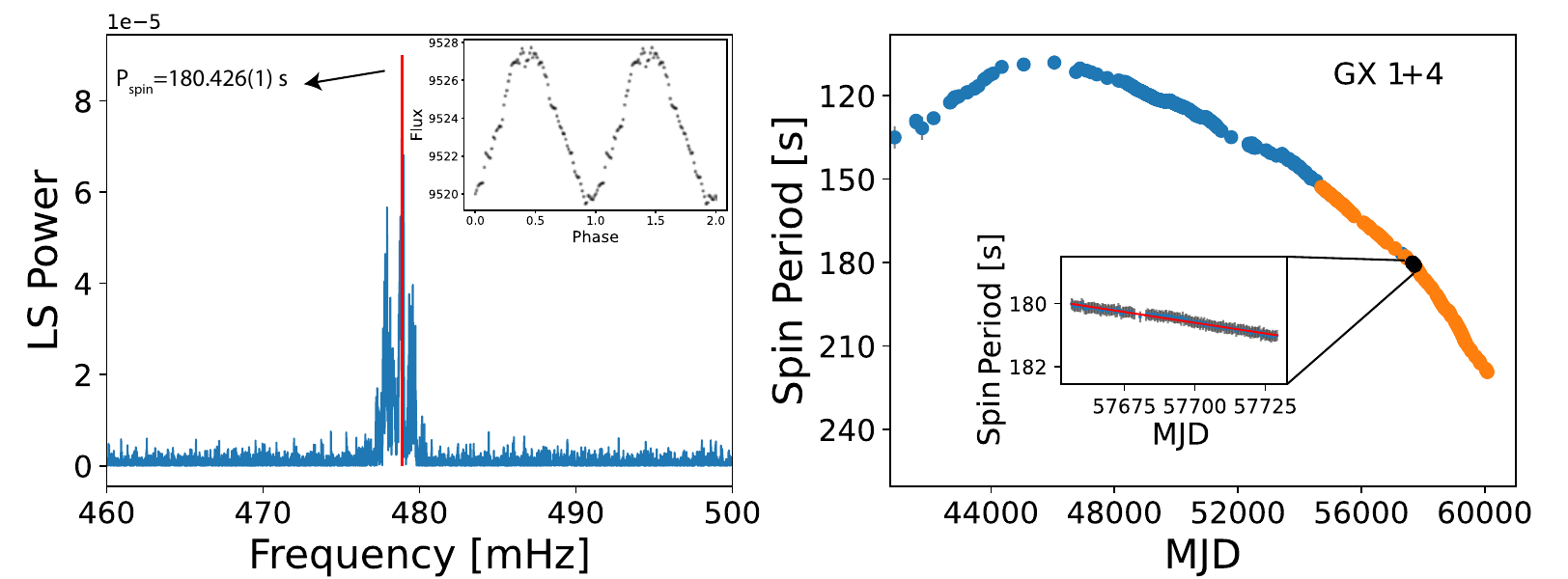}
\caption{{\it Left:} Generalized Lomb-Scargle periodogram (blue) where the frequency with the highest power is highlighted in red. This periodogram corresponds to the full {\it K2} light curve of \gx. There are other peaks around the main peak which suggests that the period has changed during the observations (see Section \ref{sec:res}). The inset shows the light curve folded at the 180.426 s period, at binned at 80 bins/cycle. {\it Right:} Historical evolution of the spin period from measurements in various instruments listed in Table B.1 of \citet{2012A&A...537A..66G}, such as {\textit RXTE} and {\textit CGRO}/{\textrm BATSE} among others (see text). {\textit Fermi}/{\textrm GBM} measurements are shown in orange symbols, while {\it K2} measurements are shown in black symbols. The inset shows the spin evolution during the {\it K2} observations reported here, where the red line shows a simple linear fit to the spin-down trend. \label{psd}}
\end{center}
\end{figure*}

The GLS periodogram from the lc1+lc2 light curve shows a highly significant peak at the period P=180.426(1) s (Figure \ref{psd}) and few other peaks close by. \citet{1997ApJ...482L.171J} reported a period of 124.17$\pm$0.04 s on 1996 April 26 in optical wavelengths. Several studies using high energy observations, previous and after the detection of the spin period in optical, already reported periods in the range of $\sim$120 to $\sim$170 seconds \citep[see][and references therein]{2012A&A...537A..66G}. Moreover, the spin period 
is known to evolve, with a spin-up phase from 1970 until 1984 and a spin-down phase since then. The most recent measurement reported from a \nustar\ observation taken on October 2015 yielded a 178.778$\pm$0.006 s spin period \citep{2018MNRAS.478..448J}. On their Figure 6, \citet{2012A&A...537A..66G} nicely shows the evolution of the spin period until 2010. I used the data from their Table B.1 and update their figure by including the optical period from \citet{1997ApJ...482L.171J}, \suzaku\ \citep{2017ApJ...838...30Y}, \nustar, {\textit Fermi} \citep{2020ApJ...896...90M} and the {\it K2} period detection reported here (see right panel in Figure \ref{psd}). Overall, the spin period found in the {\it K2} data confirms the spin-down trend determined from the other data.

The long term coverage of {\it K2} allows to search for changes in the spin period during more than 70 days. I extracted the periodograms from 1-day, consecutive and overlapping (50\% overlapping) slices of the light curve, which show strong power at the frequency of the spin of the neutron star. During consecutive slices the period increases, following the spin-down already observed in high energies (Figure \ref{psd}). The spin-down rate during the {\it K2} observations, determined by a simple linear fit, is $\sim$1.61$\times$10$^{-7}$ s/s. 




According to the orbital ephemeris from \citet{2006ApJ...641..479H}, the {\it K2} observations covered the 0.92 to 0.98 orbital phases, and in agreement with $I$-band measurements reported by \citet{2017A&A...601A.105I}, the optical emission increased toward the periastron passage. \citet{2006ApJ...641..479H} also present ephemeris for a possible eclipse of the neutron star, and the {\it K2} observations covered the phase range from 0.67 to 0.73 from the inferior conjunction. \swift/{\textrm BAT} Transient Monitor observations during the {\it K2} observations show that the source was in a low hard X-ray flux state during lc1 and part of lc2, while afterwards, the hard X-ray flux increased by a factor of about four (see Figure \ref{bat}). The \swift/{\textrm BAT} light curve shows a slow rise after about MJD 57690, reaching a maximum count rate of 0.043 c s$^{-1}$ or a flux\footnote{Following \url{http://www.dsf.unica.it/~riggio/Scripts/crab\_to\_erg.js} and \citet{2005SPIE.5898...22K}} of 2.6$\times$10$^{-9}$~\fluxcgs\ which translates into a luminosity of 5.7$\times$10$^{36}$ \lumcgs\ at a distance of 4.3 kpc \citep{2012A&A...537A..66G}. The tandem variability observed between {\textrm BAT} and {\it K2} light curves supports the scenario proposed by \citet{1997ApJ...482L.171J} where the optical light arise from reprocessed X-rays.




The current spin down rate could be caused by a retrograde rotating disk, which extracts angular momentum from the pulsar, with an increased spin down rate at higher X-ray luminosities. The analysis of the {\it K2} and \swift/{\textrm BAT} light curve does not support this scenario because an steady spin down rate is observed even after the increase of the X-ray luminosity. Moreover, as pointed out by \citet{2012A&A...537A..66G}, a retrograde disk lasting for about 40 years needs further investigation. 

As an alternative, \citet{2012A&A...537A..66G} explore the scenario of quasi-spherical accretion onto the neutron star as a possible explanation to the observed long term behavior of the spin rate. In the case that an accretion disk cannot be formed through wind accretion, depending on the source luminosity, accretion can proceed via free-fall of matter towards the magnetosphere when L$_{X}$ is above a few 10$^{36}$ \lumcgs\ while for lower luminosities, the accreting material forms a hot shell around the magnetosphere, being later accreted through instabilities in the magnetosphere. Free-fall accretion seems unlikely to have proceeded during the {\it K2} observations because of the low \swift/{\textrm BAT} luminosity (free fall accretion would require L$_{X}$ above 10$^{37}$ \lumcgs), which suggest that the settling accretion regime could have been at work. 


\subsection{\igr}

\begin{figure*}
\includegraphics[scale=0.25]{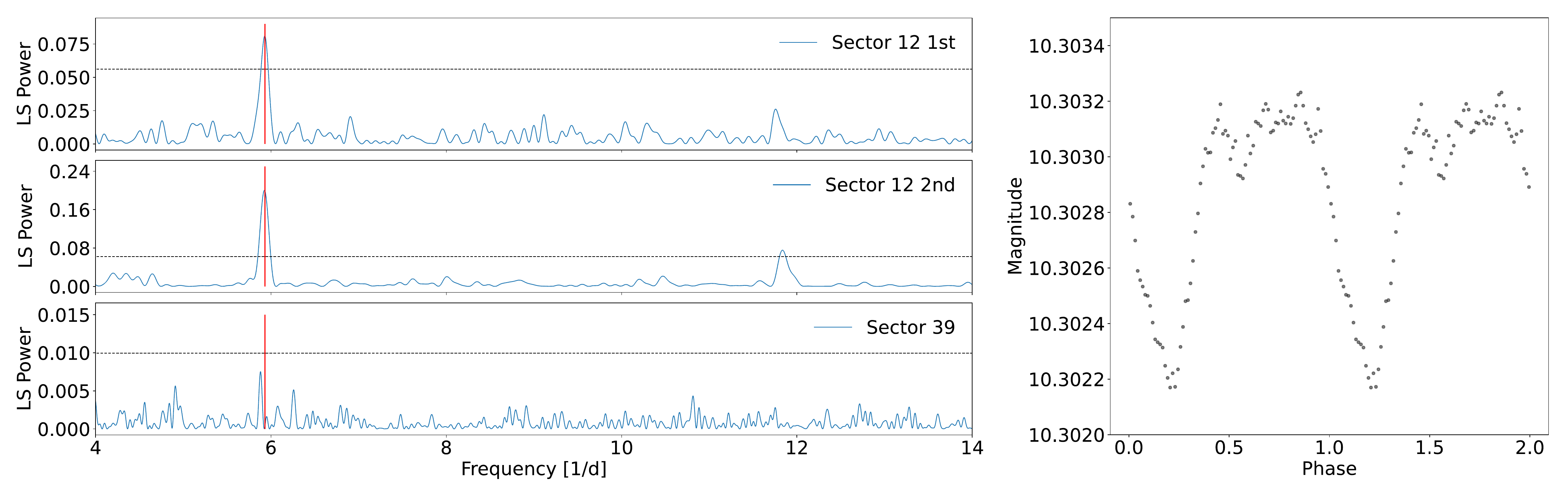}
\caption{Generalized Lomb-Scargle periodograms of the \tess\ observations of \igr\ where the frequency with the highest power is highlighted in red. These periodograms correspond to the first (top) and second (middle) portion of the Sector 12 \tess\ observations, and to Sector 39 observations (bottom). The horizontal dashed black lines represent 3 $\sigma$ detection probabilities. The right hand panel shows the light curve folded at the 242.839 min period, at binned at 80 bins/cycle}
\label{fig:igr_psd}
\end{figure*}

The GLS periodogram from the first two portions of the \tess\ light curve of \igr\ revealed a strong peak at a frequency of 5.9299 d$^{-1}$, corresponding to a period of 242.839(2) minutes (14570.34 seconds). The first harmonic of this period is also significantly detected in the power spectrum. I interpret this period as the neutron star spin period, being the first time that it is detected at any wavelength. Figure \ref{fig:igr_psd} shows the power spectrum of each portion of the \tess\ light curve. It is noticeable that the 242.839 min period is not detected during the \tess\ observations performed during sector 39, on May 2021. The 2-20 keV MAXI light curve (panel {\it c} in Fig. \ref{fig:igr_sg}) does not point to an increase/decrease of the X-ray flux during the May 2021 with respect to May 2019, which, being related with the accretion rate, could point to the origin of the non-detection of the spin of the neutron star, which remains unknown. 

\citet{2019MNRAS.485..851Y} constructed models of symbiotic X-ray binaries with various improvements over past models, such as the accretion settling regime. In their figure 2, \cite{2019MNRAS.485..851Y} presents a P$_{spin}$-L$_{X}$ diagram for different accretion scenarios (disc or wind-accretion) and different evolutionary stages of the companion (CHeB, core helium burning; EAGB, early AGB). At a distance of $\lesssim$~3.7 kpc, the X-ray luminosity of \igr\ is $\lesssim$~7$\times$10$^{34}$ \lumcgs\ \citep{2007A&A...470..331M} and with the detected spin period of 14570.34 seconds, \igr\ is located in the region of this diagram where other symbiotic X-ray binaries are found. The yet-unknown orbital period however, precludes to distinguish between the various models.

\section{Conclusions} \label{sec:conc}

By searching for the neutron stars spin period on the {\it K2} and \tess\ light curves of the X-ray symbiotics \gx\ and \igr, I have found that:

\begin{itemize}
    \item During the {\it K2} observations in 2016, the neutron star in \gx\ continued to spin down at a rate of 1.61$\times$10$^{-7}$ s/s. It is clear however, that since the beginning of the spin-down phase back in 1984 until now, the spin down rate is not constant (Fig. \ref{psd}). The \swift/{\textrm BAT} data during the same epoch shows an increase in the X-ray luminosity, which as expected if the optical light results from reprocessed X-rays, was accompanied by an optical brightening. 

    \item The increase in the X-ray luminosity during the {\it K2} observations was not high enough to change the trend of the spin period. These changes have been previously observed, in X-rays, and at higher X-ray luminosities \citep{1997ApJ...481L.101C}. 

    \item I report here, for the first time, the detection of a modulation in the \tess\ light curve of the symbiotic X-ray binary \igr . The 242.839 min period is interpreted as the period of the neutron star spin. This period is transient, detected only during the observations performed in 2019 and while absent in 2021. The non-detection of the spin period in 2021 does not seem to be related with the luminosity state of the source, given that neither the X-ray nor the optical luminosity changed significantly between the years 2019 and 2021. Further observations could elucidate the reasons behind the non-detection of the spin period. 

    \item The spin period of 242.839 min and the X-ray luminosity of about 10$^{34-35}$ \lumcgs\ \citep{2007A&A...470..331M} of \igr\ agree with model predictions and matches the location of other symbiotic X-ray binaries in the P$_{spin}$-L$_{X}$ diagram \citep{2019MNRAS.485..851Y}.

\end{itemize}

\begin{acknowledgements}
I thank the anonymous referee for useful remarks. GJML is a member of CIC-CONICET (Argentina) and acknowledge support from grant ANPCYT-PICT 0901/2017. This paper includes data collected by the Kepler mission and obtained from the MAST data archive at the Space Telescope Science Institute (STScI). Funding for the Kepler mission is provided by the NASA Science Mission Directorate. STScI is operated by the Association of Universities for Research in Astronomy, Inc., under NASA contract NAS 5–26555. This research made use of Lightkurve, a Python package for Kepler and TESS data analysis (Lightkurve Collaboration, 2018). 
\end{acknowledgements}

%
%

\bibliographystyle{aa}
\bibliography{example}

\end{document}